\documentclass[pre,twocolumn,showpacs,floatfix]{revtex4}

\usepackage{amsmath}
\usepackage{amsfonts}
\usepackage{amssymb}
\usepackage{graphicx}

\begin{document}

\title{Brownian molecular motors driven by rotation-translation coupling}
\author{Brian Geislinger}
\author{Ryoichi Kawai}
\email{kawai@uab.edu}
\affiliation{Department of Physics, University of Alabama at Birmingham,
Birmingham, AL 35294} 

\begin{abstract}
We investigated three models of Brownian motors which
convert rotational diffusion into directed translational motion by switching
on and off a potential.  In the first model a spatially
asymmetric potential generates directed translational motion by rectifying
rotational diffusion. It behaves much like a conventional
flashing ratchet.  The second model utilizes both rotational diffusion and drift
to generate translational motion without spatial asymmetry in the potential.
This second model can be driven by a combination of a Brownian motor
mechanism (diffusion driven)
or by powerstroke (drift driven) depending on the chosen parameters.  In the 
third model, elements of both the Brownian motor and powerstroke mechanisms 
are combined by switching between three distinct states. Relevance
of the model to biological motor proteins is discussed.
\end{abstract}

\maketitle

\section{Introduction}

The mechanisms of motor protein motility have been a 
major research topic in biophysics and biomedical research
for several decades~\cite{Howard2001}.  Pioneering work by Huxley and his 
co-workers~\cite{Huxley1957,Huxley1971} established a mathematical model for
muscle fiber 
contractions based on a so-called powerstroke~\cite{Howard2001,Tyska02}
involving the 
conformational change induced by ATP hydrolysis in the 
cross-bridges formed
between the myosin and actin filaments.  This 
conformational change in the neck region of myosin has been 
confirmed by various
experiments~\cite{Rayment1993,Whittaker1995,Jontes1995,%
Houdusse1999,Houdusse2000,Forkey03}.
Based on Huxley's model, 
it has been believed that this conformational change induces a 
swinging motion of the myosin 
(lever arm hypothesis)~\cite{Howard2001,Geeves02}, rotational 
motion about the neck which is in turn converted into linear 
translational motion.  This model suggests that a motor protein 
utilizing a powerstroke would behave like a stepping motor with a 
fixed step length per ATP consumption.  In addition, the step length 
is expected to be proportional to the neck length of the 
proteins.  

Recent single molecule experiments testing a 
variety of protein mutants with different neck lengths support the lever arm 
mechanism of protein 
motility~\cite{Anson1996,Uyeda1996,Sakamoto2003,Purcell2002,%
Schott2002,Ruff2001,Warshaw2000}.  Other recent experiments have emphasized both
the length of the motor protein neck as well as the magnitude of the angle
through which it
swings during a powerstroke in an attempt to explain the unusually long step
length of Myosin Id~\cite{Kohler2003}.  
However, some similar experiments showing unexpected step sizes for a given neck
length appear to be inconsistent with the lever arm
model~\cite{Ishijima1998,Itakura1993,Trybus1999,Perreault-Micale2000,%
Homma2000,Tanaka2002,Lister2004,Nishikawa2002,Rock2001,Iwane2005}.
In these experiments,
proteins must somehow move over several steps during a single chemical 
cycle in order to produce the observed step size.  
Some experiments showed that the speed of the motor protein is 
independent of the neck length in contradiction with other 
experiments~\cite{Perreault-Micale2000}.
Furthermore, the step length of certain motor 
proteins seems too long compared with their neck
length~\cite{Kitamura1999,Ali2004} and some diffusive steps appeared to
be necessary~\cite{Veigel2002,Geeves02,Vale03}.

The filaments along which motor proteins move are polymers made up of 
polar units with a ``plus'' and a ``minus'' end. Most myosin-based motor
proteins move 
preferentially towards the plus end of the actin filament track.  
Myosin VI moves in the opposite
direction~\cite{Ali2004,Amber1999,Nishikawa2002,Rock2001}. 
In addition, other motor proteins such as Myosin IXb recently have been observed
to move in the opposite direction under certain
conditions~\cite{Inoue2002,Oconnell2003}.  
These anomalous motor proteins are structurally very similar to the their 
plus-ended counterparts.  It has been observed that when the lever arm is
engineered in a certain way, the motor moved in the opposite
direction~\cite{Tsiavaliaris04}.  In the case of Myosin VI, it has been 
speculated that the naturally occurring protein moves preferentially towards the
minus end of the filament through a similar mechanism~\cite{Amber1999}. 
However, the actual mechanism determining the directionality still remains
unknown.

The majority of motor proteins are comprised of two motor domains 
that work in conjunction with each other to produce directed motion.  
The two heads are particularly crucial to the models put forth to 
describe processive motors such as Myosin V, which move through 
many ATP hydrolysis cycles before detaching from 
its filament~\cite{Howard2001,Leibler1993}.  Most 
of the prevailing models on processive motor movement require the 
two motor domains to work in sequence,
alternating attachment states 
allowing the motor to move forward without completely releasing from 
the filament~\cite{Yildiz2003,Vale03,Tyska03,Vilfan05}.  However, some motor
proteins such as Myosin IXb have only a single
motor domain and yet is still a processive 
motor~\cite{Inoue2002,Post2002,Oconnell2003}.  In addition, numerous other
experiments have 
been performed on single headed molecular motors, both naturally 
occurring and derived single-headed mutants.  They show a continued ability
to move and produce a force against an external load despite the obvious
structural deficiencies with regards to the models based on two-head
mechanisms~\cite{Kitamura1999,Yanagida2002,Watanabe2004,Veigel2005,Purcell2005}.

Apart from these recent experimental findings, new models for 
motor proteins based on the idea of Brownian motors have 
been intensively investigated during the last
decade~\cite{magnasco93,Astumian1994,Astumian1997,Prost1997,Prost1995}.  
When Brownian particles are in a periodic but spatially 
asymmetric potential and away
from thermal equilibrium, they can generate mechanical work by rectifying
thermal fluctuation~\cite{Reimann2002}. Since motor proteins move along an 
asymmetric filament and ATP hydrolysis creates nonequilibrium 
conditions, it has been proposed that motor 
proteins could be described by a Brownian motor 
model~\cite{magnasco93,Astumian1994,Astumian1997,Prost1997,Prost1995}.  
A Brownian 
motor based on the flashing ratchet model moves as a 
step motor in consistent with experiments.  
However, Brownian motors also provide for the possibility 
of moving more than one step per chemical cycle in contrast 
to the powerstroke model.  Furthermore, the step length of 
the Brownian motor is independent on the neck length.  These features 
possibly account for some of the experimental data which the 
powerstroke model fails to explain~\cite{Kitamura1999,Terada2002}.  

The powerstroke model and Brownian motor model has been 
thought of as two opposing extremes in the search for motility mechanisms 
of molecular motors.  In the former, thermal fluctuations have 
no role at all.  On the other hand, the latter is incapable 
of producing motion without thermal fluctuations.  As a result, it was 
commonly believed that only one of them is the correct mechanism.  
However, it has recently been postulated that both mechanisms 
may be involved in the actual motor proteins~\cite{Veigel2002,Geeves02}.  

The purpose 
of the present work is to develop a single mathematical model that 
includes both powerstroke and Brownian motor mechanisms.  Since the 
coupling between the rotational and linear translational degrees of 
freedom is essential to the powerstroke mechanism, our first step 
is to develop a model which couples rotational motion to linear 
motion (Section~\ref{sec:model}).  In the first model 
(Section~\ref{sec:freerotor}), we demonstrate that free thermal diffusion 
in the rotational degree of freedom is sufficient to induce 
directed linear motion via a flashing ratchet mechanism.  A 
model based on the behavior of motor proteins that is discussed in 
Section~\ref{sec:tethered} shows directed
translational motion driven by either a Brownian motor or a 
powerstroke mechanism depending on the parameters used.  A more realistic motor
protein 
model involving three chemical states is proposed in 
section~\ref{sec:tristate}.  This motor is
simultaneously capable of both powerstroke and Brownian motor mechanisms during
a
single chemical cycle.  Relevance of the present models to actual biological
motors is discussed in the final section.

\section{Rotation-Translation Coupling}
\label{sec:model}

\begin{figure}
\centering\includegraphics[width=8.6cm]{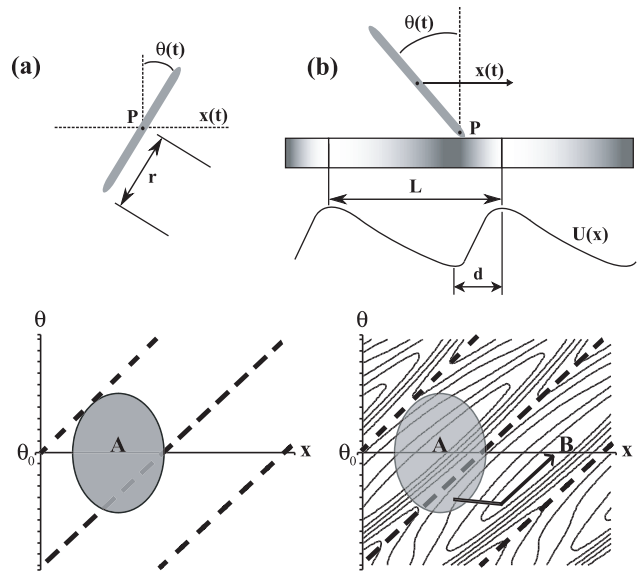}
\caption{Model for molecular motors with a coupled rotational and
translational mechanism for producing directed motion.  (a) \textit{Unbound
state}.  Unattached of its filament track, the motor is free to thermally
diffuse in both $x$ and $\theta$.  The resulting probability distribution $A$
is shown in the lower image.  In principle, $D_x \neq D_\theta$ so $A$ is
depicted as an ellipse.  The dashed lines depict the location of the potential
barriers in the bound state for a point of reference.  (b) \textit{Bound state}.
 Upon binding to its filament, the pivot point $P$ of the motor shifts towards
the binding site, activating the coupling between rotation and translation.  The
potential contours shown in the lower image corresponding to the bound state,
depict the probability of remaining in the same potential well at $A$ or moving
to the next at $B$.  As can be seen here, an asymmetric potential $U(x)$ is
required to bias the motion of the motor.}
\label{fig:model}
\end{figure}

While various mathematical models for molecular motors driven by
rotation-translation coupling have been proposed in the
past\cite{li98,porto01}, these models are not closely related to known biological
molecular motors.  A more realistic model is necessary to explain the 
experimentally observed behaviors of biological molecular motors, such as step
size distributions and velocity-force relationships.  Here, we try to develop a
mathematically simple, yet sufficiently realistic model by taking into
account experimental observations as much as possible.

Motor proteins undergo two major conformational changes during a chemical cycle,
first when ATP hydrolysis takes place and second when ADP is released from the
protein.  The main change occurs between the head and neck of the protein such
that the angle between the neck with respect to the head changes.  For
simplicity, we consider a simple rod of length $\ell=2r$ as a motor particle
whose
degrees of freedom are defined as the position of the center of mass $x$ and a
rotational angle $\theta$ around the
center (measured from the normal to the filament). Since free rotation does not
generate translational motion, it is therefore necessary to
couple the rotational motion with the translational degree of freedom.  Since
the release of ADP occurs when the protein is attached to the filament, a large 
conformational change in the neck associated with the release would produce a
rotation about the point of attachment to the filament.  The resulting motion
about that pivot point produces translational motion of the center of the
motor that is proportional to $r \sin\theta$ as
shown in
Fig.~\ref{fig:model}.

The potential energy of the motor in the bound state is determined by
the location of 
the contact point $x_P$ and the orientation of the motor $\theta$.  In
general, these two degrees of freedom are weakly coupled but not in a way to
produce a
translational motion out of the swing motion. Therefore, we assume
an uncoupled form of potential energy:
\begin{equation}
V(x_P,\theta) = U_{rot}(\theta) + U_{trans}(x_P)
\end{equation}
where $U_{rot}$ represents a conformational change in the neck region of the
protein and $U_{trans}$ is a
binding energy between the motor and the filament.
Since the filament is periodic, the binding energy is also 
periodic as $U_{trans}(x_P+L) = U_{trans}(x_P)$ where $L$ is the
period of the filament. Noting that $x_p$ and center-of-mass coordinate $x$
are related by $x_P=x-r \sin\theta \approx x -
r\theta$~\cite{note1}, the binding energy as a
function of
$x$ and $\theta$ is given by
\begin{equation}
V(x,\theta) = U_{rot}(\theta) + U_{trans}(x-r\theta)
\label{eqn:u}
\end{equation}
This potential energy function
provides the desirable form of coupling between
rotational and translational degrees of freedom, allowing for the conversion of
rotational motion to translational motion as shown below.

\section{A two-state Brownian ratchet driven by rotational diffusion}
\label{sec:freerotor}

Although motor proteins in general have several chemical states
during ATP hydrolysis, we will restrict this discussion to a two-state model
describing a bound and unbound state for a motor protein.  When a potential is
absent (unbound state)
the motor freely diffuses in its two degrees of freedom, $x$ and $\theta$. As
soon as the potential is turned on (bound state), the motor drifts toward
a local potential minimum. Switching between these two states will produce net
movement in the $x$ direction when the potential is asymmetric along $x$. 
To illustrate this ratchet mechanism, consider
motors confined to an equilibrium point $A$ of the potential well, as seen in 
Fig.~\ref{fig:model}(b). When the potential is removed, the motors diffuse over
the shaded
area in
Fig.~\ref{fig:model}(a). As the potential turns on again, a portion of the
shaded area now belongs to the basin of next potential minimum $B$ in
Fig.~\ref{fig:model}(b).  The motors that reached this area advance to the new
minimum $B$.  The rest of the motors move back to the original minimum $A$.  In
the particular case shown in Fig.~\ref{fig:model}, the translational diffusion
alone is too small to generate directed motion.  However, the addition of the
rotational diffusion brings the motors to the next basin.

This two-dimensional Brownian ratchet mechanism can be put into a
simple mathematical expression in a slow switching limit~\cite{Astumian1994}.  
Assuming that the potential $U_{trans}$ is sufficiently deep and the
transitions 
between unbound and bound states are slow, the initial probability distribution
of the motors is
given by $P(x,\theta,t=0)=\delta(x-x_A)\delta(\theta-\theta_0)$.
When the potential is removed, the distribution spreads with diffusion
constants $D_x$ and $D_\theta$ as
\begin{eqnarray}
& &P(x,\theta,t) = P_{rot}(\theta,t)P_{trans}(x,t)  \nonumber  \\
& &             = \frac{1}{2\pi t \sqrt{D_x D_\theta}} \exp \left
(-\frac{(x-x_a)^2}{2D_x t} - \frac{(\theta-\theta_0)^2}{2D_\theta t} \right ).
\label{eq:P}
\end{eqnarray}
When the potential is reinstated at $t=\tau$, the probability that the
motors leave the basin of the 
current potential wells in the forward (+x) and backward (-x) direction are
given by
\begin{eqnarray}
W_+(\tau) &=& \int_{-\infty}^{\infty} dx \int_{-\infty}^{(x-d)/r} d\theta
P(x,\theta,\tau) \nonumber  \\
&=& \frac{1}{2} \text{erfc} \left ( \frac{d}{\sqrt{(2D_x + 2 r^2
D_\theta)\tau}} \right ),
\label{eqn:forwardW}
\end{eqnarray}
and
\begin{eqnarray}
W_-(\tau) &=& \int_{-\infty}^{\infty} dx \int_{-(x+L-d)/r}^{+\infty} d\theta
P(x,\theta,\tau) \nonumber  \\
&=& \frac{1}{2}\text{erfc} \left ( \frac{L-d}{\sqrt{(2D_x + 2 r^2
D_\theta)\tau}}\right ),
\label{eqn:backwardW}
\end{eqnarray}
where $d$ is the shortest distance between the potential minimum and maximum,
[see Fig. 1(b)] and erfc is the complementary error function. If
the motor is in the unbound state for period $t_u$ and in the bound state for
$t_b$ during one on-off cycle, then the average velocity of the motor is about
\begin{equation}
\overline{v} \approx \frac{L}{t_b+t_u} (W_+(t_u) - W_-(t_u)).
\label{eqn:adiabaticspeed}
\end{equation}
Here we neglected the possibility for the motors to diffuse beyond the nearest
neighbor potential wells.  When the potential is symmetric ($d=L/2$), $W_+=W_-$
and thus no net directed motion occurs. 

\begin{figure}
	\centering
	\includegraphics[width=8.6cm]{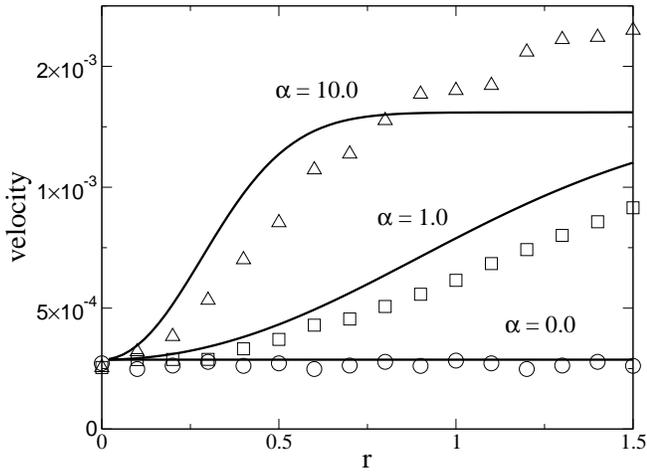}
	\caption{
The relationship between average motor velocity and the motor radius $r$.  The circle markers correspond to $\alpha$=0.0, the squares to $\alpha$=1.0, and the triangles to $\alpha$=10.0.  The case $\alpha$=0.0 corresponds to a purely translational ratchet.  For nonzero $\alpha$ with a sufficiently large value for $r$, the rotational component of the motion provides a unique contribution to the motor velocity.  The solid lines correspond to the long time limit velocity obtained from Eq.~\ref{eqn:adiabaticspeed}.  Parameter values:
$t_u$=$5.0$, $t_{b}$=$100.0$, $K$=1.0, $D$=$0.0075$, $d$=$0.36$.}
	\label{fig:adiabatic}
\end{figure}

The above discussion can be further confirmed by a direct simulation of
Langevin equations for the overdamped motor operated at a temperature $T$:
\begin{subequations}
\label{eqn:langevin}
\begin{align}
\gamma_x \frac{dx}{dt} &= -\frac{\partial V(x,\theta)}{\partial x}\sigma(t)
+ \sqrt{2 k_B T \gamma_x }~\xi_x(t) + F \\
\gamma_\theta \frac{d\theta}{dt} &= -\frac{\partial V(x,\theta)}{\partial
\theta}\sigma(t) + \sqrt{2 k_B T \gamma_\theta}~\xi_\theta(t),
\end{align}
\end{subequations}
where $F$ is an external load and 
$\xi_i$ is a Gaussian white noise define
by
\begin{equation}
\langle \xi_i(t) \xi_j(t') \rangle = \delta_{ij} \delta(t-t') \qquad i,j \in (x,
\theta).
\end{equation}
The on-off function $\sigma(t)$ takes values 0 and 1 alternatively with
transition
rates $k_{on}$ and $k_{off}$. 
Although in general $\gamma_x$ would have an angular dependency according to the
current orientation of the rod, here we assume that it
takes constant values. 

As a simple model system we use a ratchet potential in the $x$
direction:
\begin{equation}
\label{eqn:Ux}
U_{trans}(x) = U_0 \left [ \frac{3}{4}\cos\left ( \frac{2\pi}{L} x\right ) -
\frac{1}{4}\sin\left(\frac{4\pi}{L} x\right ) \right ].
\end{equation}
and a simple harmonic potential with a spring constant $K$ for rotation:
\begin{equation}
U_{rot}(\theta) = \frac{K}{2} \left ( \theta - \theta_0 \right )^2.
\end{equation}
For simplicity, a value of $K$=1.0 will be used throughout the current study.

To reduce the number of parameters, we normalize time, distance and energy as
\begin{equation}
\tilde{t} \equiv \frac{U_0 t}{L^2 \gamma_x}, \quad \tilde{x} \equiv \frac{x}{L},
\quad \tilde{U} \equiv \frac{U}{U_0}
\end{equation}
and accordingly parameters are normalized as $\tilde{r} = r/L$, $\tilde{K} =
K/U_0$, $\tilde{F} = FL/U_0$.
For simplicity, we will omit the tilde on the normalized quantities.
Then, the equations of motion (\ref{eqn:langevin}) become
\begin{subequations}\label{eqn:langevin_normal}
\begin{align}
\frac{dx}{dt} &= -\frac{\partial V(x,\theta)}{\partial x}
\sigma(t) + \sqrt{2 D}~\xi_x + F\\
\frac{d\theta}{dt} &= -\alpha\frac{\partial V(x,\theta)}{\partial
\theta}\sigma(t)  + \sqrt{2 \alpha D}~\xi_\theta
\end{align}
\end{subequations}
where the diffusion constant is defined by $D=k_B T/U_0$.  A dimensionless
constant $\alpha = r^2 \gamma_x/\gamma_\theta$ is determined by the detailed
shape of the motor. The contribution of rotational degree of freedom is
controlled by the shape of the motor and hence the value of $\alpha$.
When $\alpha=0$, the rotational motion has no contribution and the
present model is equivalent to a conventional flashing ratchet. For
$\alpha>1$,
the rotational diffusion plays  a more significant role than the translational
diffusion.

\begin{figure}
	\centering
	\includegraphics[width=8.6cm]{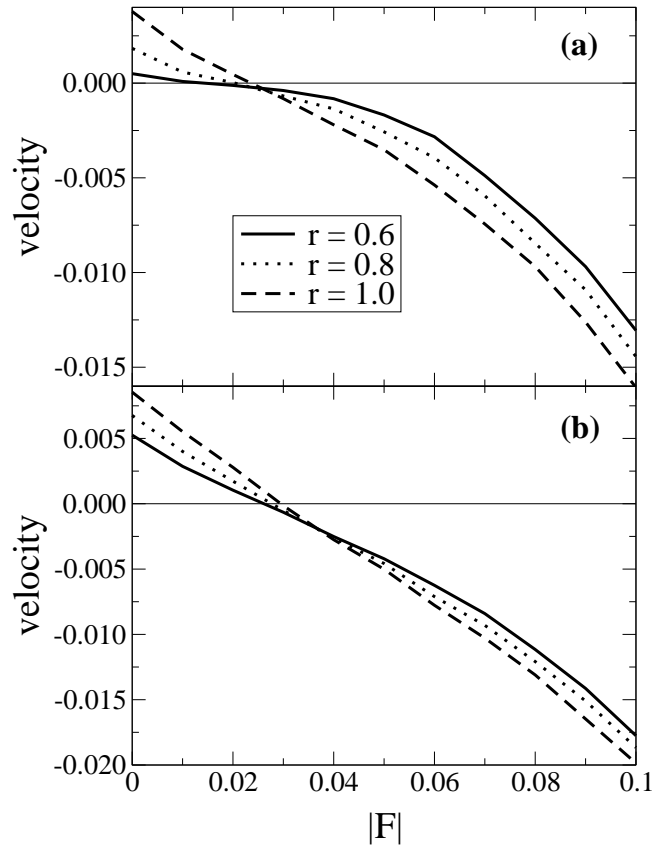}
	\caption{The effect of applying a constant load $F<0$ to a motor
according to
the equations of motion given by Eq.~(\ref{eqn:langevin_normal}).  (a) We see a
distinct nonlinear behavior towards the application of an increasing external
load $F$ when the motion relies primarily upon the rotational ratchet mechanism.
 Parameter
values:  $\alpha$=$10.0$; $K$=1.0; $k_{on}$=$k_{off}$=$0.5$.  
(b) The nonlinear behavior disappears when equal proportions of both
translational and rotational ratchet are used.  Parameter values: 
$\alpha$=$1.0$; $K$=1.0; $k_{on}$=$k_{off}$=$0.5$.}
	\label{fig:externalforce}
\end{figure}

Integrating the coupled Langevin equations in Eq.~(\ref{eqn:langevin_normal})
with the Heun method, we can explore the dynamics of this stochastic model. 
Examining the trajectories of numerous motors allows us to determine the average
behavior of the system.  For our purposes, an average over 500 realizations of
the system will be sufficiently accurate.

Figure~\ref{fig:adiabatic} shows that the velocity contains no dependence on $r$
for $\alpha$=0.0 as expected.  For nonzero values of $\alpha$, the velocity
begins to increase with the motor length $r$ beyond the values expected from the
translational ratchet alone, showing that the rotational component of the motion
introduces an additional mechanism for producing directed motion in the motor. 
The speed of the motor increases in a roughly linear fashion with respect to the
motor length $r$ similar to the way that most motor proteins are thought to
behave.  As the length of the motor increases, a change in the angle $\theta$
will give the motor a longer reach, enabling it to move several potential
periods over one cycle with a sufficiently long motor.  As seen in
Fig.~\ref{fig:adiabatic}, the rotational motion contributes to the motion with
increasing $r$ until the motor is long enough to reach the nearest neighboring
potential well.  Beyond a certain length, the velocity is expected to saturate,
until it is then long enough to reach the next nearest neighbor.  This behavior
is corroborated by the expected behavior of the long time limit case for these various
values of $\alpha$.  Figure~\ref{fig:adiabatic} shows close agreement between
Eq.~\ref{eqn:adiabaticspeed} described above and the stochastic simulations.

By applying a constant load $F<0$, we
can investigate the ability of this model to do work against a load.  Figure
\ref{fig:externalforce} shows the results of two cases: $\alpha$=$10.0$ where
the rotational diffusion dominates
and $\alpha$=$1.0$ where rotational and translational diffusion equally
contribute to the motion.  The limit where the motor moves primarily under
rotational diffusion exhibits a nonlinear response to
an applied load.  Upon moving towards the limit of equal contributions
(\textit{i.e.} $\alpha$=$1.0$) the nonlinearity disappears.   The nonlinearity
is therefore due to the anisotropic diffusion.

\section{A two-state Brownian motor with powerstroke}
\label{sec:tethered}

\begin{figure}
	\centering
	\includegraphics[width=8.6cm]{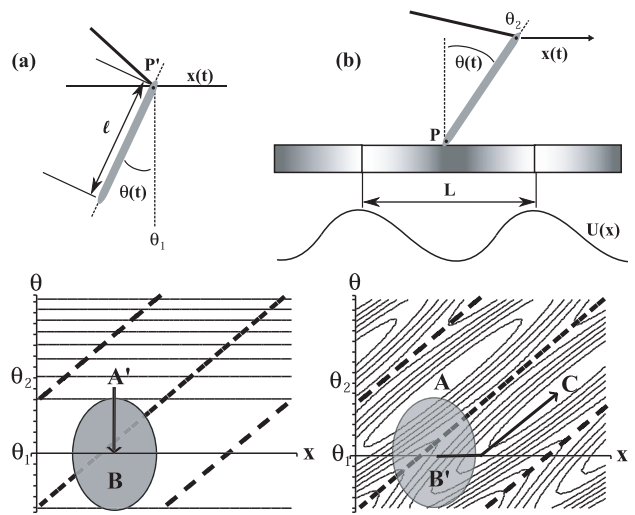}
	\caption{A model for motor proteins with one end of a motor of length
$\ell$
tethered to an external cargo.  (a) \textit{Unbound state}.  As before, the
motor is free to diffuse in $x$ and $\theta$, but $\theta$ drifts toward a stable angle
$\theta_1$.  Overlaid on the potential contours is the new probability distribution at $B$ centered about 
$\theta_1$, away from the potential minimum of the bound state (shown at $A'$). 
(b) \textit{Bound state}.  On binding, the pivot $P$ shifts, giving
rise to the coupling term in $U_{trans}(x)$.  Here the stable angle for the
bound state
$\theta_2$ is not necessarily equal to $\theta_1$.  Given a large enough change
in the protein conformation between the bound and unbound states
$\ell\Delta\theta$=$\ell(\theta_2 - \theta_1)$, motion can occur
deterministically without the need for the probability distribution at $B$.}
	\label{fig:tether}
\end{figure}

In the previous model, coupling between the rotational and translational degree of
freedom takes place only when the motor protein is bound to an active site on
the filament.
The rotation
toward an equilibrium angle advances the motor to a minimum within the basin of
the potential well. We shall call this rotation the \textit{forward} stroke.  
The forward stroke however cannot drive the motor over the potential
barrier and only diffusion, rotational or translational, brings the
motor to the next potential well. On the contrary, in the standard powerstroke
model
conformational change of the motor protein drives the
motor without diffusion.  The key point of the powerstroke model is
the presence of the \textit{recovery} stroke which the motor performs when it is
unbound.
Molecular motors typically consist of a motor domain attached to a neck and tail
region.  The tail region typically binds to some sort of cellular cargo or, in
the case of skeletal muscle, the tail binds together with other proteins to form
a
bundle
that works in tandem.  As opposed to our free floating motor presented in the
previous section,
the motor protein does not freely rotate in the unbound state and instead is
driven
to an equilibrium angle by the recovery stroke. 

As an alternate model to our previous Brownian ratchet, we consider a  motor
which is
not entirely free to move, but is tethered by one end to some external object
assumed to be much larger than itself.   When the motor is unbound, it rotates
about the pivot point $P'$  toward an
equilibrium angle
$\theta_1$ (recovery stroke) as illustrated in in Fig.~\ref{fig:tether}(a). 
When the motor binds to the
filament, the pivot
point shifts to the other end of the motor, $P$ in
Fig.~\ref{fig:tether}(b). The
motor rotates to a new equilibrium angle $\theta_2$ about this point (forward
stroke).   The two different equilibrium angles $\theta_1$ and $\theta_2$
represent
conformational changes in the motor proteins due to ATP hydrolysis.

Suppose that the motor is located at a bound state $A$ in
Fig.~\ref{fig:tether}(b).
Upon detaching from the filament, the recovery stroke takes the motor from $A'$
to $B$ in Fig.~\ref{fig:tether}(a).  If the conformational change,
$\ell\Delta\theta = \ell(\theta_2-\theta_1$) is large enough, the point $B$ is
in the
basin of the next potential well.  When it binds to the filament, the forward
stroke bring the motor to $C$, generating a net translational motion from $A$ to
$C$.
This process is entirely deterministic and the directed motion is generated
simply by
alternating forward and recovery rotations without diffusion.  However, the
motor
is still subject to large thermal fluctuation, which may hamper the
powerstroke.  Indeed, the thermal fluctuation brings some motors back to
the basin of the original potential well as illustrated in
Fig.~\ref{fig:tether} and thus reduces the average speed.
On the other hand, the thermal fluctuations also make a positive contribution
to the
directed motion especially when the powerstroke fails.  For example, one could
imagine a motor prematurely binding to the filament before the recovery stroke
completes.  In a deterministic model that powerstroke would fail to advance
the motor to the next potential well. 
With the addition of the thermal diffusion, some of the motors that would fail
the powerstroke are able to reach the next potential well with help of that
diffusion.  Furthermore, when the conformational change $\ell \Delta \theta$ is
too small to reach the next minimum, the Brownian
motion is still capable of producing directed motion on average.

Unlike the previous model, rotational diffusion in the unbound state is
accompanied by a drift due to the conformational change $\ell\Delta\theta$.
This rotational drift determines the direction of motion.  When the sign of
$\ell\Delta \theta$ is reversed, the motor moves in the opposite direction. 
Recent
experiments have exhibited such flux reversal for motors modified to swing
their neck in the opposite direction to the original
motors~\cite{Amber1999,Tsiavaliaris04}.
Since the symmetry is broken by these conformational changes, it is not
necessary for the potential to break the symmetry along the $x$ axis.   A simple
periodic potential in the normalized units described earlier will suffice for
our purposes:
\begin{equation}
\label{eqn:u_tether}
U_{trans}(x) = \cos(2\pi x).
\end{equation}
The net potential for the unbound and bound states are respectively given by
\begin{subequations}\label{eqn:v12t}
\begin{align}
V_1(x,\theta) &= \frac{K}{2}(\theta - \theta_1)^2  \\
V_2(x,\theta) &= \frac{K}{2}(\theta - \theta_2)^2 + U_{trans}(x-\ell\theta),
\end{align}
\end{subequations}
where $x$ is now the position defined by the neck of the motor protein at
$P'$.
The movement of the motors is determined by the Langevin equations in
normalized units:
\begin{subequations}
\begin{align}
\frac{dx}{dt} &= -\frac{\partial V(x,\theta,t)}{\partial x}
 + \sqrt{2 D}~\xi_x + F\\
\frac{d\theta}{dt} &= -\alpha\frac{\partial V(x,\theta,t)}{\partial
\theta} + \sqrt{2 \alpha D}~\xi_\theta
\end{align}
\end{subequations}
where the time-dependent potential is defined as
$V(x,\theta,t)=V_1(x,\theta)[1-\sigma(t)] + V_2(x,\theta)\sigma(t)$.


\begin{figure}
	\centering
	\includegraphics[width=8.6cm]{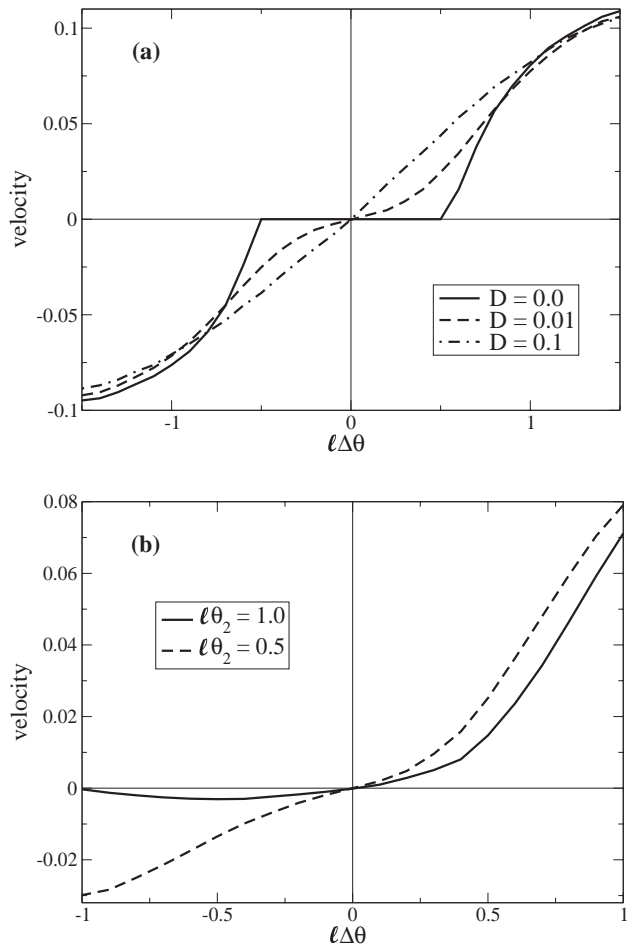}
	\caption{
(a) The average velocity of the motor $\langle \dot{x} \rangle$ as a function of
the change in the stable conformations $\ell\Delta\theta$ between the two states
with the unbound angle $\theta_{1}$ fixed.  Parameter values: $t_u$=$t_b$=$4.0$;
$K$=1.0; $\theta_{1}$=$0.0$; $\alpha$=$1.0$.  (b) The average velocity as a
function of
$\Delta\theta$ with the bound stable angle $\theta_{2}$ fixed.  Parameter
values:  $t_u$=$t_b$=$4.0$; $K$=1.0; $D$=$0.01$; $\alpha$=$1.0$.}
	\label{fig:tether_sim}
\end{figure}

\begin{figure}
	\centering
	\includegraphics[width=8.6cm]{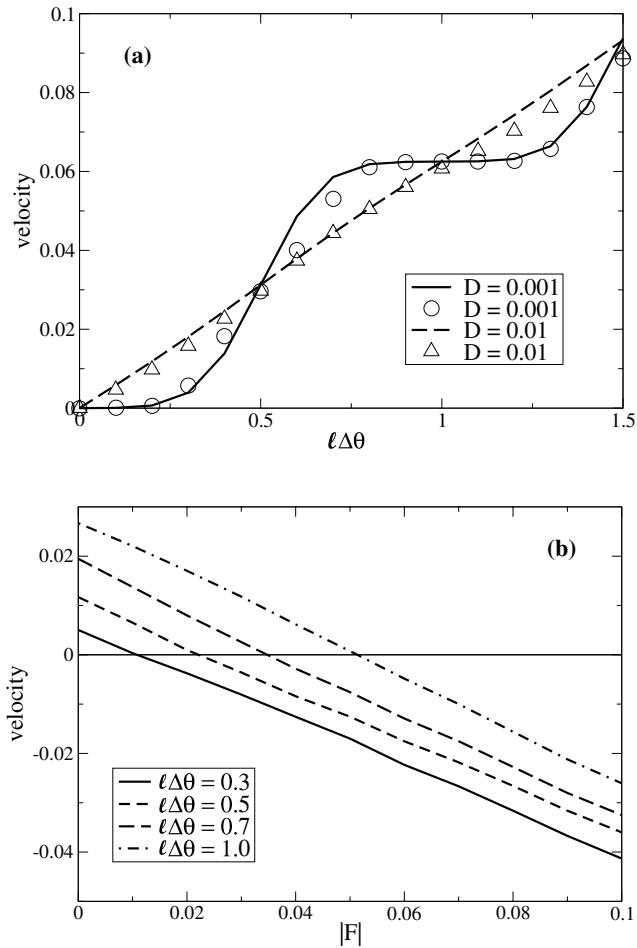}
	\caption{
(a) An analysis of the tether motor system showing both the case of
small and large values of $D$.  The markers correspond to the Langevin simulation and the lines correspond to the long time limit theory.  Parameter values: $t_u$=$t_b$=$16.0$; $K$=1.0; $\alpha$=$1.0$.  
(b) An external load $F<0$ applied to a 2 state tethered motor system. 
Parameter values: $t_u$=$t_b$=$16.0$; $K$=1.0; $D$=$0.01$; $\alpha$=$1.0$.}
	\label{fig:tether_force}
\end{figure}

We begin by examining the relative change in the conformation $\ell\Delta\theta$
between the two states $V_1$ and $V_2$.  Figure \ref{fig:tether_sim} shows the
average velocity of the motor as $\ell\Delta\theta$ is increased for several
different values of the diffusion constant $D$.  In the case of $D$=0.0, the
thermal fluctuations are disabled.  For a small neck length $\ell$ or a small $\Delta\theta$, no velocity is observed.  However,
once the critical value of $\ell\Delta\theta \approx 0.5$ is reached, the motor
begins to move without the need of the thermal fluctuations.  This deterministic
motion corresponds to a powerstoke motion, related both to the neck length of
the motor given by $\ell$ and the angle through which the motor rotates between
the unbound and bound states.  Although in principle the motor moves forward one
potential through each hydrolysis cycle, close to the critical value of
$\ell\Delta\theta$ the motor would require a very long switching time to relax
to the potential minimum of the bound state in order to achieve that expected
velocity.  

In the case of $D \neq 0$, Fig.~\ref{fig:tether_sim} shows that even for small
values of $D$ motion can be seen in the region of $\ell\Delta\theta < 0.5$ where
the deterministic powerstroke was not possible.  The thermal fluctuations of the
medium surrounding the motor can be rectified to produce a nonzero average
velocity.  The mechanism for rectifying the thermal fluctuation follows the
Brownian ratchet mechanism from the coupled rotational and translational
coupling described in Section~\ref{sec:freerotor}.  

Also of interest is the case where $\ell\Delta\theta < 0$.  This parameter
regime would be equivalent to a motor taking a power stroke in a negative
direction, and as a results Fig.~\ref{fig:tether_sim} shows an average negative
velocity for the motor.  This observation is of potential interest in describing
the motion of motor proteins such as Myosin VI, observed to move in the opposite
direction as the plus-ended Myosin V proteins.  Using the angle $\theta_1$=$0.0$
for the unbound configuration gives a completely symmetric response given the
symmetric nature of $U_{trans}(x)$.  On the other hand, by fixing the bound
stable angle $\theta_{2}$ to some non-zero value, the symmetry disappear.  To
meet the conditions for $\ell\Delta\theta <0$ with $\theta_2$ fixed, $\theta_1$
must become increasingly large to obtain any negative motion.  The harmonic
potential term in Eq.~(\ref{eqn:v12t}) quickly becomes too large to sustain
negative motion and drives the motor back towards the positive.  It is likely that this case is not well represented by the current model and this behavior is an artifact of the approximations that $U_{rot}$ is a harmonic potenial and $r\sin\theta \approx r\theta$.

In summary, the diffusionless curve in Fig.~\ref{fig:tether_sim} shows the
contribution of the deterministic motion.  When the $\ell\Delta\theta$ becomes
very large ($\ell\Delta\theta > 1.0$), the deterministic motion dominates the
model.  For small changes in the angle ($\ell\Delta\theta < 0.5$), the Brownian
ratchet mechanism dominates.  Between these two limits a combination of the two
mechanisms contribute to the average velocity of the motor.  The resulting model
forms a very robust motor that can produce directed motion under a wide variety
of parameters.

As before, the ratcheting motion found in this model can be studied analytically
in the long time limit as described before with little difference.  However the
limits of the integral in Eq.~(\ref{eqn:forwardW}) change to take into account
the change the angle $\theta_i$ between states.  The new forms for $W_+$ and
$W_-$ can be obtained by replacing $d$ with $L/2 - \ell\Delta\theta$ in
Eqs.~(\ref{eqn:forwardW}) and (\ref{eqn:backwardW}).  A nonzero
$\ell\Delta\theta$ breaks the symmetry of the system and gives rise to  $W_+
\neq W_-$ even though we are using a symmetric potential.  The resulting current
then follows the same form as stated in Eq.~(\ref{eqn:adiabaticspeed}).  In
comparing the results of the long time limit calculation to the stochastic simulations
in Fig.~\ref{fig:tether_force}(a), we again find close agreement to the motion seen
in the long time limit.

Motor proteins typically bind a large cargo to their tail in order to transport
it from one part of a cell to another.  As before, we can apply an external load
$F < 0$ to these motors to study their ability to pull such a load that might be
attached to it.  In Fig.~\ref{fig:tether_force}(b), the same type of linear
behavior is observed as before with the previous model.  However,
Figure~\ref{fig:tether_force}(b) also shows that as the value of $\ell\Delta\theta$
is increased the maximum force with which the motor pulls increases.  It has been postulated that a Brownian ratchet
is unlikely to be a mechanism for motor proteins as those models tend to produce
forces too small to account for the forces observed experimentally
\cite{Howard2001}.  Based on the present results, it appears that as we increase
the contribution of powerstroke to the function of the motor,
the  ability of the motor to transport cargo more effectively also increases.

\section{A three state model}
\label{sec:tristate}

\begin{figure}
	\centering
	\includegraphics[width=8.6cm]{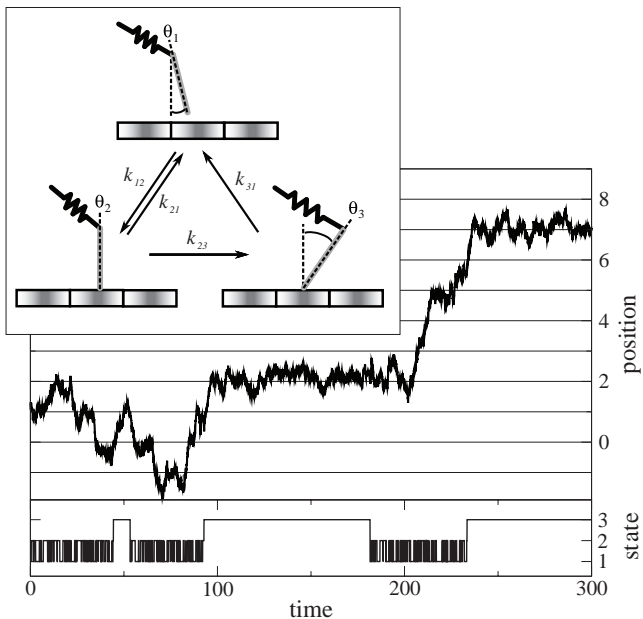}
	\caption{ A three state reaction scheme for a motor protein and a
typical trajectory for such a motor.  Rate constants governing the reaction are
labeled as $k_{ij}$ and the conformational angle for each state is $\theta_i$
where $i,j \in (1,2,3)$.  Trajectory parameter values: $D$=0.01; $\alpha$=$1.0$;
$k_{12}$=0.5, $k_{21}$=0.2, $k_{23}$=0.2, $k_{31}$=0.004; $U_2$=0.2, $U_3$=1.0; 
$K$=1.0; $\ell\Delta\theta_{12}$=0.2, $\ell\Delta\theta_{23}$=1.0.}
	\label{fig:3state_reaction}
\end{figure}

The cyclic chemical reaction that myosin undergoes as it hydrolyzes ATP can be
expressed with varying degrees of precision involving numerous substeps of the
reaction~\cite{Howard2001,Keller2000,Astumian2002a}.  
Detailed analysis of the structure of
myosin S1 has indicated the existence of at least three distinct conformational
states dependent on the state of the bound ATP
molecule.~\cite{Rayment1993,Whittaker1995,Jontes1995,Houdusse1999,Houdusse2000,
Coureux2004}.  In addition, some myosin working strokes have been resolved into
at least two distinct  substeps~\cite{Veigel1999,Veigel2002,Lister2004},
yielding more evidence of multiple conformations of the motor protein.  Here, we
will utilize three distinct conformational states with a chemical reaction
scheme similar to the simple mechanochemical trigger model put forth by Keller
and Bustamante~\cite{Keller2000}.  This setup enables us to create a hybrid
motor that contains elements of both a Brownian motor as well as a deterministic
powerstroke.

Following the cyclic reaction scheme shown in Fig.~\ref{fig:3state_reaction},
the transition between a motor containing ATP in an unbound state and its
subsequent weakly bound state to an actin filament as the ATP is broken down
into ADP and P$_i$ is reversible.  This transition induces a small
conformational change $\ell\Delta\theta_{12}=\ell(\theta_2 - \theta_1)$ which is
not big enough to perform a powerstroke.  However a back and forth transition
between this unbound and weakly bound state can generate directed motion by
rectifying rotational diffusion as discussed in the previous sections.  The
transition from a weakly bound to a strongly bound state involves the
irreversible release of the inorganic phosphate P$_i$ and is accompanied by a
larger change in the conformational state $\ell\Delta\theta_{23}=\ell(\theta_3 -
\theta_2)$ providing the powerstroke.  Completing the cycle involves the release
of ADP and binding of a new ATP molecule, returning the motor protein to an
unbound state.  Cycling along these three states, the motor has the capability
of utilizing both the Brownian motor and powerstroke mechanisms.

Comparing the typical trajectory of this model shown in
Fig.~\ref{fig:3state_reaction} to single molecule experiments performed in
recent years shows a possible mechanism explaining observed behavior.  The model
presented here exhibits occasional backward steps and multiple forward steps
during the Brownian ratchet phase similar to the events observed by Yanagida's
group~\cite{Kitamura1999,Ishijima1998}.  The additional powerstroke as
the motor enters the strongly bound state provides another step forward along
the potential, similar to the two stage working stroke observed in some
experiments~\cite{Veigel1999,Veigel2002,Lister2004}.

\begin{figure}
	\centering
	\includegraphics[width=8.6cm]{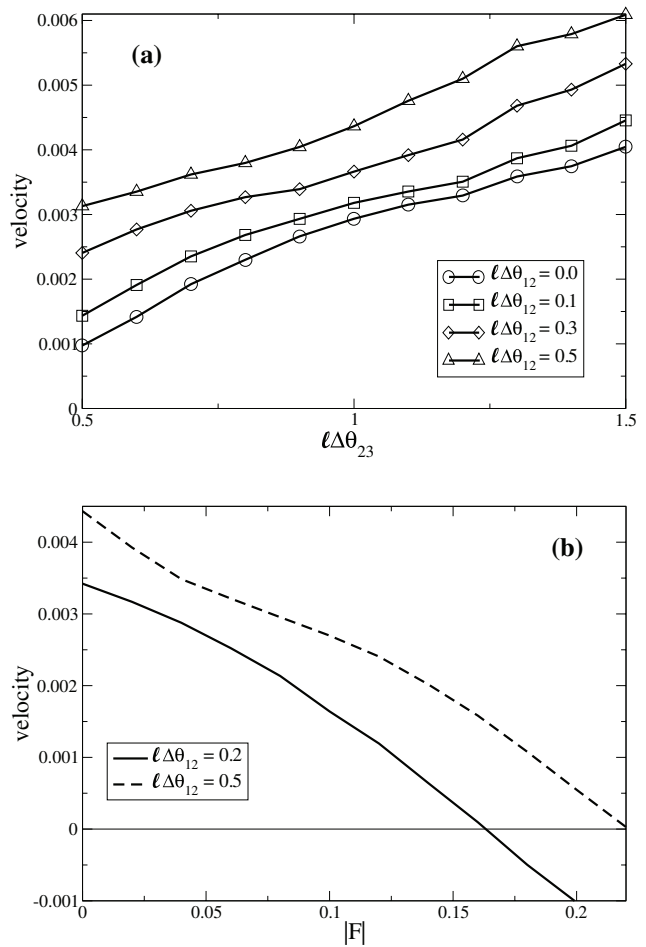}
	\caption{
(a) Motor velocity as a function of varying conformational changes from state 1
to 2, and from state 2 to 3.  Parameter values: $D$=0.01; $\alpha$=$1.0$;
$k_{12}$=0.5, $k_{21}$=0.2, $k_{23}$=0.2, $k_{31}$=0.004; $U_2$=0.2, $U_3$=1.0; 
$K$=1.0.  
(b) Motor velocity as a function of an applied load $F < 0$.  Parameter values:
as in (a) with $\ell\Delta\theta_{23}$=1.0. }
	\label{fig:3state}
\end{figure}

Figure~\ref{fig:3state}(a) demonstrates the change in velocity as both the
magnitude of the powerstroke and the conformational change associated with the
Brownian motor are increased.  In the case of $\ell\Delta\theta_{12}$=0, the
Brownian motor mechanism makes no contribution to the velocity of the motor.  As
$\ell\Delta\theta_{12}$ is increased, the Brownian motor mechanism has the
capability of tripling the overall speed of a motor, implying that the motor
could take on average two extra steps during the Brownian motor phase of the
hydrolysis cycle with adequate tuning of the parameters.  As seen
in Fig.~\ref{fig:3state}(b), this three state system has the added advantage of
withstanding a larger external load opposing the motion than the simpler two
state model. 

\section{Discussion}

Conventional Brownian ratchets encounter various difficulties when applied to
biological motors.  In the present model, 
coupling the translational and
rotational degrees of freedom preserves the conventional
translational ratchet mechanism, but also provides a means to produce
translational motion through rectifying thermal fluctuations in the rotational
degree of freedom. Further, with the addition of an asymmetric conformational
change 
between bound and unbound states, we obtain a model capable of rectifying
thermal fluctuations without the need for an asymmetric potential.  In
the conventional Brownian ratchet model the direction of the motion is
determined by a spatial asymmetry in the translational degree of freedom. 
However, in the present model, the direction of motion is determined by the
asymmetry in the conformational change in the motor proteins rather than the
asymmetry in the filament, consistent with experimental
observation~\cite{Amber1999,Tsiavaliaris04}.
If this
conformational change between the bound and unbound states is large enough, the
motor gains the added ability to move by powerstroke without diffusion. 
Depending on the parameters
chosen, either mechanism can be utilized ultimately gaining the benefits of
both.

With the minimal three
chemical state description detailed in the previous section, we introduce a
realistic model for single headed motor proteins, which captures various
qualitative properties of motor proteins. 
Recent experiments on monomeric Myosin I have emphasized the importance of both
the neck length and the degree of neck rotation in determining the step size of
a motor protein~\cite{Kohler2003}.  The present model takes both factors into
account in defining the coordinate of conformational change as
$\ell\Delta\theta$.  Other single molecule experiments have resolved at least
two distinct conformational changes during the working stroke of Myosin
I~\cite{Veigel1999}, one corresponding to a displacement of $\sim 6.5\, nm$ and
another of $\sim 5.5\, nm$, both very close to the size of the actin filament
repeat.  The multiple steps observed in Myosin I can be explained by the
different phases of motion through an ATPase cycle of the present model.  As
seen in the trajectory in Fig.~\ref{fig:3state_reaction}, a diffusive step
followed by a powerstroke would each produce displacement in a series of steps
equal to the filament periodicity $L$.

Many single molecule experiments with Myosin II have observed step sizes on the
order of
$4-10\, nm$~\cite{Molloy1995,Mehta1997,Ishijima1998,Tanaka1998,Kitamura1999,
Tyska1999,Kad2003}.  However, according to Tanaka {\it et
al.}~\cite{Tanaka1998}, a random orientation of myosin with respect to an actin
filament produces the observed average working stroke of $\sim 5.5\, nm$.  When
myosin is close to its natural orientation with respect to actin as it would be
in a whole fiber, the step size is closer to three times that value, or
$\sim 15\, nm$. Kitamura {\it et al.}~\cite{Kitamura1999} observed substeps of
$\sim 5.3\, nm$ along actin subunits.  As the dwell time of these substeps had
no dependence on ATP concentration, it was proposed that an average of 2.5
substeps during a single ATPase cycle makes up the observed $\sim 13\, nm$ step
per ATP molecule consumed.  A simple deterministic motor protein should take a
single step per ATP molecule consumed.  However, as illustrated in
Fig.~\ref{fig:3state}(a), a motor protein with $\ell\Delta\theta_{12}$=0.5 on
average should take roughly two 5-$nm$ diffusive steps along the filament
subunits in addition to the powerstroke during the consumption of a single ATP
for a total step size of $15\, nm$.  

In the case of dimeric, processive motor proteins like Myosin V, models for the
processivity involve some type of coordination between the two motor domains of
the myosin.  
Investigations into single-headed Myosin V motors have revealed a 
directional force dependence on the kinetics that regulate ATP 
hydrolysis~\cite{Purcell2005,Veigel2005}.
These force dependent kinetics indicate a possible method of communication 
between the two individual 
motor domains.  When the leading head of the motor pair finds its binding 
site on the actin filament and enters a strongly bound state,
the strain caused
on the trailing head encourages its release from the filament.  At the same time
the release of the leading head is slowed, effectively anchoring the motor
protein to the
filament allowing the trailing head to swing forward and find its next binding
site.
While this motion of the trailing head forward is thought as a
powerstroke process~\cite{Forkey03,Yildiz2003,Vale03,Tyska03,Vilfan05},
some recent experiments suggest an alternative.  The
working stroke of a single-headed Myosin V motor protein produces a displacement
of
only $\sim 25\, nm$, significantly shorter than the observed step size of
$36\, nm$ for wild type two-headed Myosin V~\cite{Veigel2002}.  The motion of
this wild type
Myosin V has since been resolved into a $12\, nm$ step followed by a $24\, nm$
step~\cite{Uemura2004}.  Due to the insufficient length of the working 
stroke of a single-headed Myosin V motor,
it has been proposed that the powerstroke of the leading head partially
positions the trailing head and the remaining distance is covered by a
diffusive process~\cite{Veigel2002,Geeves02}. Others have shown that single
headed Myosin V molecules can move processively with the motor
protein held in close proximity to the filament~\cite{Watanabe2004}.  In this
study it was 
proposed that through a series of diffusive steps utilizing a strain sensors
similar to the one demonstrated by Veigel {\it et al.}~\cite{Veigel2005} and
Purcell {\it et al.}~\cite{Purcell2005}, a single motor protein can achieve a
step size of $32\, nm$.  

Myosin VI, another dimeric motor protein, has a neck
length much less than that of Myosin V, yet still produces a step length of
similar size~\cite{Ali2004,Lister2004}.  As a result, any deterministic
powerstroke of this motor protein probably has little to do with its step and is
most likely driven by thermal motion.  The present model provides
a single mechanism allowing a motor to take advantage of both diffusion
as well as a subsequent powerstroke/anchoring. Given the large neck of
Myosin V, a significant powerstroke on the order of $25\, nm$ would be
reasonable based on a large value for $\ell$, followed by a Brownian motor phase
positioning the trailing head at its full step distance from the leading head.  
On the other hand, a model for Myosin VI could rely heavily upon the 
Brownian motor mechanism in order to reproduce the observed behavior.  
We will further explore these dimeric motor proteins based on our model in a
future paper~\cite{dimerpaper}.

Some monomeric motor proteins such as Myosin IX are thought to involve a
diffusive step in order to achieve some measure of
processivity~\cite{Post2002,Inoue2002,Nalavadi2005,Kambara2006}.  However, to
maintain this processivity monomeric motor proteins must remain in the vicinity
of a filament while in a diffusive state.  Both Nalavadi
{\it et al.}~\cite{Nalavadi2005} and Kambara {\it et al.}~\cite{Kambara2006}
have proposed a mechanism.  The
structure of Myosin IX contains a unique insertion in the motor domain that may
harbor a second binding site.  This insertion could tether the protein to its
filament while still allowing enough freedom for the protein to diffuse.  The
motor protein then diffuses towards its next binding site before entering
another powerstroke. Given the assumption of this special tether in Myosin IX,
the present model provides a mechanism to efficiently guide the motor to its
next binding site by rectifying thermal fluctuations before entering a strongly
bound state.

All of these single molecule constructs have been observed to produce a force on
the order of a few
$pN$~\cite{Howard2001,Molloy1995,Purcell2005,Veigel2005,Watanabe2004,Tyska1999,
Mehta1999}.  Examining the normalized load $F$ from
Eq.(\ref{eqn:langevin_normal}), for $L$=5.4 $nm$, the approximate length of an
actin subunit, and $U_0$=20$k_BT$  one unit of normalized force is about
$15.8\, pN$.  The maximum forces generated by the monomers shown in
Fig.~\ref{fig:3state}(b) are then approximately $2.6-3.5\, pN$, comparable to
the forces observed experimentally.

In conclusion, we have developed a realistic model of motor proteins which
simultaneously includes the traditional Brownian ratchet model and the
powerstroke model in a simple unified form.  The properties of this model
are qualitatively consistent with recent experimental results.  
Quantitative investigation is currently underway.  Furthermore, we are
applying the present model to dimeric processive motors as well as muscle fiber.
Although the model is developed for the myosin superfamily, we
believe it can be applied to other molecular motors such as kinesin.

\section{Acknowledgments}
The authors would like to acknowledge Kimmie Farris and Erin Darnell for their
help in this work.  This work was partially supported by National Science
Foundation (NSF) GK-12 Award No. 0139108.

\bibliographystyle{apsrev}

\begin{thebibliography}{70}
\bibitem[{not()}]{note1}
\bibinfo{note}{Here we assumed $\theta << 1$ just for symplicity. Although
  actual angle may not be small, the main conclusions in the present work are
  valid without this approximation.}
\end{thebibliography}

\end{document}